\documentclass[fleqn,usenatbib]{mnras}

\usepackage{newtxtext,newtxmath}

\usepackage[T1]{fontenc}

\DeclareRobustCommand{\VAN}[3]{#2}
\let\VANthebibliography\thebibliography
\def\thebibliography{\DeclareRobustCommand{\VAN}[3]{##3}\VANthebibliography}

\usepackage{graphicx}	
\usepackage{amsmath}	

\title[EGMF constraints: 145 months of \textit{Fermi}-LAT data]{Constraints on the extragalactic magnetic field strength \\ from blazar spectra based on 145 months of \textit{Fermi}-LAT observations}
\author[Podlesnyi, Dzhatdoev, and Galkin]{
E. I. Podlesnyi,$^{1,2,3}$\thanks{E-mail: podlesnyi.ei14@physics.msu.ru}
T. A. Dzhatdoev,$^{3,2}$
V. I. Galkin$^{1,2}$
\\
$^{1}$Department of Physics, Federal State Budget Educational Institution of Higher Education M.V. Lomonosov Moscow State University,\\
1(2), Leninskie Gory, GSP-1, 119991 Moscow, Russia\\
$^{2}$Skobeltsyn Institute of Nuclear Physics (SINP MSU), Federal State Budget Educational Institution of Higher Education M.V. Lomonosov Moscow State University,\\
1(2), Leninskie Gory, GSP-1, 119991 Moscow, Russia\\
$^{3}$Institute for Nuclear Research of the Russian Academy of Sciences,\\
7a, 60th October Anniversary Prospect, 117312 Moscow, Russia
}

\date{Accepted XXX. Received YYY; in original form ZZZ}

\pubyear{2022}

\graphicspath{{./}{figures/}}

\begin{document}
\label{firstpage}
\pagerange{\pageref{firstpage}--\pageref{lastpage}}
\maketitle

\begin{abstract}
Properties of the extragalactic magnetic field (EGMF) outside of clusters and filaments of the large-scale structure are essentially unknown. The EGMF could be probed with $\gamma$-ray observations of distant (redshift $z > 0.1$) blazars. TeV $\gamma$ rays from these sources are strongly absorbed on extragalactic background light photons; secondary electrons and positrons produce cascade $\gamma$ rays with the observable flux dependent on EGMF parameters. We put constraints on the EGMF strength using 145 months of \textit{Fermi}-LAT observations of the blazars 1ES 1218+304, 1ES 1101-232, and 1ES 0347-121, and imaging atmospheric Cherenkov telescope observations of the same sources. We perform a series of full direct Monte Carlo simulations of intergalactic electromagnetic cascades with the \texttt{ELMAG 3.01} code and construct a model of the observable spectra inside the point spread functions of the observing instruments for a range of EGMF strengths. We compare the observed spectra with the models for various values of the EGMF strength $B$ and calculate the exclusion statistical significance for every value of $B$. We find that the values of the EGMF strength $B \le 10^{-17}$ G are excluded at a high level of the statistical significance $Z > 4 \sigma$ for all the four options of the intrinsic spectral shape considered (power-law, power-law with exponential cutoff, log-parabola, log-parabola with exponential cutoff). The value of $B = 10^{-16}$ G is not excluded; it is still a viable option of the EGMF strength. These results were obtained for the case of steady sources.
\end{abstract}



\section{Introduction} \label{sec:intro}
\begin{table*}
\centering
\begin{tabular}{l|c|c|c|c}
Source name & 4FGL catalog name & $z$; reference & IACT observational period(s) & Reference(s)\\
\hline
{1ES 1218+304} & {4FGL J1221.3+3010} & 0.184; (1) & 2012--2013 & \citet{Madhavan2013}\\
{1ES 1101-232} & {4FGL J1103.6-2329} & 0.186; (2) & 2004--2005 & \citet{Aharonian2006, Aharonian2007a}\\
{1ES 0347-121} & {4FGL J0349.4-1159} & 0.188; (3) & Aug.--Dec. 2006 & \citet{Aharonian2007b}\\
\end{tabular}
\caption{List of considered blazars, their 4FGL catalog \citep{Abdollahi2020} names, cosmological redshifts, IACT observational periods and corresponding references. References for $z$: (1): \citet{SDSS2017, SDSS2021}; (2): \citet{Remillard1989}; (3): \citet{Woo2005}.}
\label{tab:sources}
\end{table*}

The strength of the extragalactic magnetic field (EGMF) $B$ in voids of the large-scale structure (LSS) of the Universe remains weakly constrained. Upper limits on $B$ obtained with the Faraday rotation measures method for the EGMF correlation length $\lambda = 1$ Mpc are $\sim1$ nG \citep{Pshirkov2016}. In principle, much weaker values of $B$ in voids are viable; therefore, a qualitatively different approach could be required in order to measure these weak magnetic fields. For reviews on various theoretical EGMF models the reader is referred
to \citet{Grasso2001}, Sect. 2--4, 6 of \citet{Durrer2013}, and to \citet{Vachaspati2021}; compilations of existing attempts on constraining the EGMF parameters and discussions of related phenomenology could be found in Sect. 5 of \citet{Durrer2013} and in \citet{Han2017,Dzhatdoev2018,AlvesBatista2021}.

Observations of extragalactic sources in the high energy (HE, $E > 100$~MeV) and very high energy (VHE, $E > 100$~GeV) $\gamma$-ray domains allow to probe the EGMF with $B$ ranging from $10^{-20}$ G to $10^{-12}$ G. The basic idea of this approach is known at least from 1989 \citep{Honda1989}: primary $\gamma$ rays emitted by the source are partially absorbed on extragalactic photon fields by means of the pair production (PP) process ($\gamma\gamma \rightarrow e^{+}e^{-}$); secondary electrons and positrons (hereafter called ``electrons'' for simplicity) get deflected by the EGMF and then produce secondary (cascade) $\gamma$ rays by means of the inverse Compton (IC) process ($e^{-} \gamma \rightarrow e^{-'} \gamma^{'}$ or $e^{+} \gamma \rightarrow e^{+'} \gamma^{'}$) so that the energy, angular, and temporal characteristics of the observable $\gamma$-ray emission appear to be sensitive to the EGMF strength and structure.

Many changes to and improvements over the original implementation of this approach by \citet{Honda1989} naturally came in over the intervening years. Most notably, the extragalactic HE $\gamma$-ray sky turned out to be dominated by distant ($L > 100$ Mpc) blazars \citep{Abdollahi2020,Wakely2008} --- active galactic nuclei with relativistic jets pointing towards the observer. The maximal energy of $\gamma$ rays detected from these blazars is so far only 20--30 TeV (e.g., \citet{Aharonian1999}); therefore, it is very important to account for the PP process on extragalactic background light (EBL) photons \citep{Nikishov1962,Gould1967} dominating at the energy $E < 100$ TeV and not only the PP process on cosmic microwave background light (CMB) photons \citep{Jelley1966}.

In practice, obtaining constraints on the EGMF strength and correlation length usually requires the observation of the same sources with two different kinds of instruments: 1) imaging atmospheric Cherenkov telescopes (IACTs) such as H.E.S.S. \citep{Hinton2004, Giavitto2018}, MAGIC \citep{Aleksic2016a,Aleksic2016b}, VERITAS \citep{Krennrich2004, Park2016} and 2) space $\gamma$-ray telescopes such as \textit{Fermi}-LAT \citep{Atwood2009}. The majority of these studies rely solely on the energy spectra; the results are often reported for $\lambda = 1$ Mpc. We note that the EGMF could be inhomogeneous and different for the directions to various sources.

\citet{Neronov2010}, using approximately one year of \textit{Fermi}-LAT observations, assuming power-law with exponential cutoff primary $\gamma$-ray spectra (see Eq.~(\ref{eq:PLExp})) and steady sources find that in the case of $B < 3 \cdot 10^{-16}$ G cascade $\gamma$ rays would overshoot the upper limits on the \textit{Fermi}-LAT spectra, implying the lower limit $B \geq 3 \cdot 10^{-16}$ G. Relaxing the source stability assumption, \citet{Dermer2011} obtain another, more conservative lower limit $B \geq 10^{-18}$ G. For some blazars, two different, but reasonably good fits to the measured $\gamma$-ray spectra are possible: 1)~the one with a significant (frequently dominating) contribution of the cascade component to the total $\gamma$-ray intensity at the energy $E < 100$ GeV (``maximal cascade model''), 2) the opposite case of the ``minimal cascade model'' \citep{Taylor2011}. Naturally, the assumed primary $\gamma$-ray spectra are also different for these two cases. \citet{Taylor2011} find $B \geq 10^{-17}$ G assuming that the sources were active on a timescale $T > 3$ years (the duration of the \textit{Fermi}-LAT observations then available). The results of \citet{Vovk2012} are similar to those of \citet{Taylor2011}.

\citet{Finke2015}, using 70 months of \textit{Fermi}-LAT data and considering various EBL models, obtain $B \gtrsim 10^{-19}$~G. However, \citet{Arlen2014} argue that even the hypothesis of $B = 0$ cannot be ruled out given the uncertainties of the EBL models, blazars' intrinsic spectra of primary $\gamma$ rays, jet opening angle and beaming pattern, and jet viewing angle. We note that this last result was obtained with the use of 42 months of \textit{Fermi}-LAT data.

Small ($\delta \ll 1$ rad) typical deflection of the cascade electrons corresponds to the magnetically broadened cascade (MBC) regime \citep{Abramowski2014}, as opposed to the pair halo (PH) regime for $\delta > 1$ rad \citep{Aharonian1994}. A search for the MBC pattern in \textit{Fermi}-LAT data was conducted in \citet{Abramowski2014}. The negative result of this search allows to exclude $B = 3 \cdot 10^{-16}- 3 \cdot 10^{-15}$ G. \citet{Archambault2017}, using VERITAS data on the blazar 1ES 1218+304, exclude a range of EGMF strengths around $B = 10^{-14}$~G at the 95\% confidence level through non-detection of the MBC.

Observations of transient events such as gamma-ray bursts (GRBs) allow to conduct a search for pair echoes delayed with respect to the primary $\gamma$ rays \citep{Plaga1995,Ichiki2008,Murase2008,Veres2017}. However, even for the case of a very bright GRB 190114C the sensitivity of \textit{Fermi}-LAT is not sufficient to obtain any constraints on the EGMF parameters \citep{Dzhatdoev2020}.

Finally, \citet{Ackermann2018} (hereafter A18), using 90 months of \textit{Fermi}-LAT data and making a number of assumptions (in particular, that ``accounting for the cascade contribution does not change the best-ﬁt spectrum of the central point source in the entire \textit{Fermi}-LAT energy band by more than 5$\sigma$'') obtain the lower limit $B > 3\cdot10^{-13}$~G for $T > 10^{7}$ years and $B > 3\cdot10^{-16}$~G for $T > 10$ years. The first of these results is in stark contrast with many of the works mentioned earlier, e.g. \citet{Taylor2011,Vovk2012,Finke2015}. Therefore, it is worth considering how the assumptions made in A18 (in particular, the one mentioned above) could have influenced the constraint on $B$.

In the present work, we put constraints on the EGMF strength, dispensing with any assumptions about the contribution of the cascade component to the observable $\gamma$-ray spectrum. As a dataset, we use the spectra of three blazars detected with IACTs and \textit{Fermi}-LAT, namely 1ES 1218+304, 1ES 1101-232, and 1ES 0347-121 (see Table~\ref{tab:sources}). All three sources have close values of the cosmological redshift $z$. This allows us to use the same simulations (see Sect.~\ref{sec:modelling}) for all the sources.

In Sect.~\ref{sec:Fermi-LAT} we describe the \textit{Fermi}-LAT data analysis. In Sect.~\ref{sec:modelling} we outline the simulation procedure allowing to calculate the observable $\gamma$-ray spectrum for various values of the EGMF strength. The statistical analysis method utilized in this work is presented in Sect.~\ref{sec:statistics}. The main results of our work are described in Sect.~\ref{sec:results}. In Sect.~\ref{sec:discussion} we discuss various systematic effects that could influence the EGMF constraints; as well, we outline prospects of EGMF measurements with next-generation $\gamma$-ray observatories. Finally, we conclude in Sect.~\ref{sec:conclusions}.

\begin{figure*}
\includegraphics[width=0.99\textwidth]{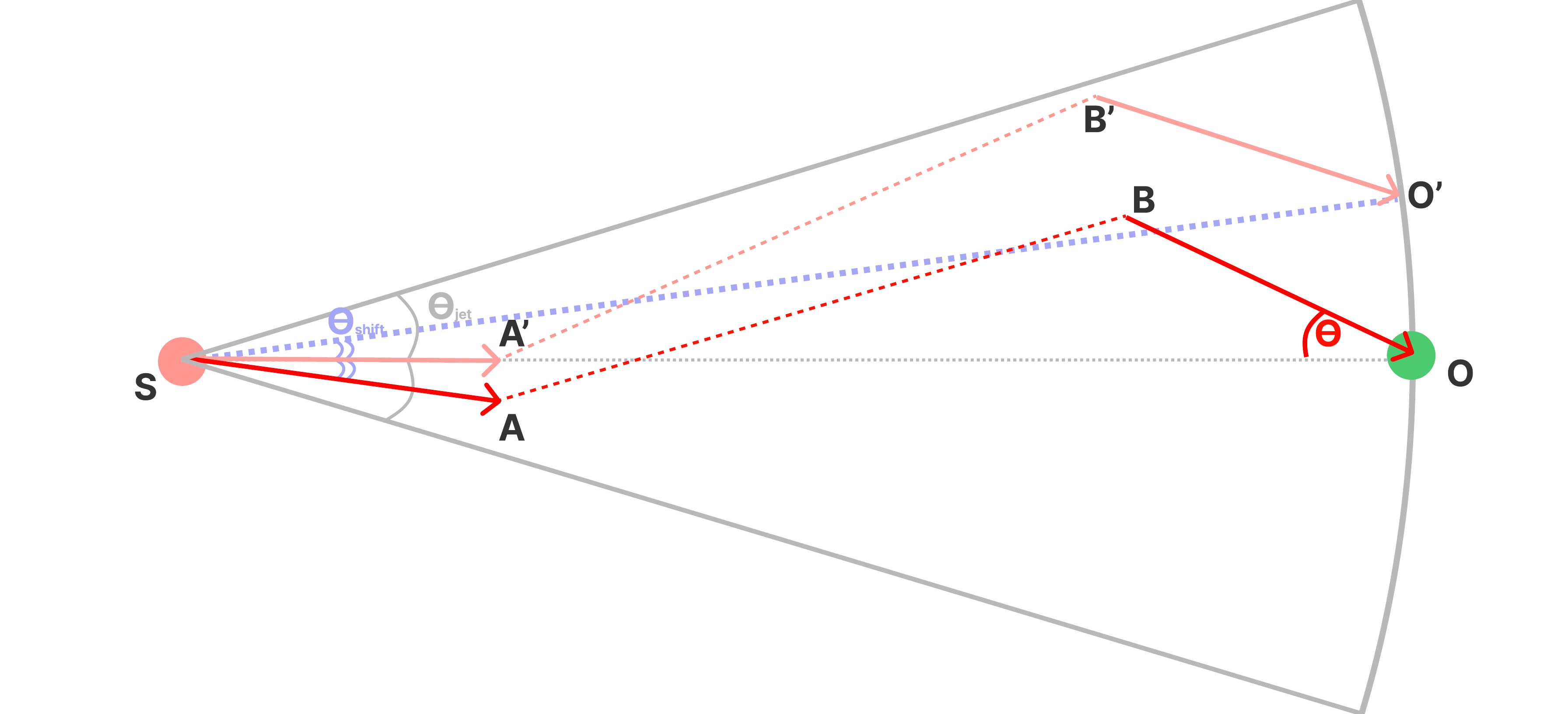}
\caption{A sketch of the geometry of intergalactic electromagnetic cascade (not to scale). \label{fig:scheme}}
\end{figure*}
\section{\textit{Fermi}-LAT data analysis} \label{sec:Fermi-LAT}

We reconstruct the spectral energy distributions (SEDs = $E^{2}dN/dE$) for the sources listed in Table~\ref{tab:sources} using publicly-available \textit{Fermi}-LAT \citep{Atwood2009} data. The datasets were extracted from the LAT Data Server of Goddard Space Flight Center for the observation dates between 2008-08-04 and 2020-09-20 (in total approximately 145 months of live data). The regions of interest (ROI) are squares with the width of 10$^{\circ}$ and the centers at the source positions taken from the 4FGL catalog. We have applied the energy selection from 1 GeV to 500 GeV.

The \texttt{Fermitools} package available at \url{https://github.com/Fermi-LAT/Fermitools-conda} (version 1.2.23) as well as the \texttt{fermiPy} package from \url{https://github.com/fermiPy/fermipy} (version 0.19.0) were used for the data processing. Namely, a search for the maximum of binned likelihood was carried out for each source with the following parameter values: \texttt{zmax} = 90, \texttt{evclass}  = 128, \texttt{evtype} = 3, \texttt{edisp} = True, instrument response function (IRF) = \texttt{P8R3\_SOURCE\_V2}. Since A18 did not detect any angular extension(s) for the considered $\gamma$-ray sources beyond those caused by instrumental effects, we assume that the sources are point-like.

All nearby sources from the 4FGL \textit{Fermi}-LAT source catalog within $10^{\circ}$-square with the center at the target position were accounted for in the analysis, as well as both galactic and isotropic $\gamma$-ray backgrounds according to the \texttt{gll\_iem\_v07} and \texttt{iso\_P8R3\_SOURCE\_V2\_v1} models, respectively. The normalization of sources within $1^{\circ}$ from the target was left free with their spectral shapes fixed, and all values of spectral parameters of sources beyond the $1^{\circ}$-circle from the ROI center were fixed to their 4FGL catalog values. After the ROI fitting the SEDs of target sources were obtained with the \texttt{fermipy.GTAnalysis.sed} method with all its parameters set to their default values.
        
We considered energy binning with 4 and 8 bins per decade (bpd) but in the latter case the SEDs revealed too many upper-limit bins at higher energies so we adopted the former option. Finally, we performed a similar analysis for the same sources in the energy range of 100 MeV -- 500 GeV for the observation dates between 2008-08-04 and 2021-01-14 but even for the 2 bpd binning the analysis yielded only upper limits for 1ES 1101-232 and 1ES 0347-121 below 1 GeV; therefore, in what follows we adopt the energy threshold of 1 GeV.

\section{Modelling extragalactic electromagnetic cascades and observable $\gamma$-ray spectrum} \label{sec:modelling}

To simulate the development of intergalactic electromagnetic cascades in the magnetized expanding Universe, we use the open-source Monte Carlo code \texttt{ELMAG 3.01} \citep{Blytt2020, Kachelries2012} in the full three-dimensional particle propagation mode.

A simplified scheme of the relevant geometry is shown in Fig.~\ref{fig:scheme}. Let us consider a conical blazar jet (with the half-opening angle $\theta_{\mathrm{jet}}$) launched by the source (situated at $S$). The observer is situated at $O$. Consider a primary VHE $\gamma$ ray emitted by the source (primary $\gamma$ ray direction $SA$), absorbing on a EBL/CMB photon initiating an intergalactic electromagnetic cascade so that an observable $\gamma$ ray ($BO$) is registered.

The \texttt{ELMAG 3.01} code allows to obtain the energy-angular distribution of these observable $\gamma$ ray as follows. The code actually propagates primary $\gamma$ rays in the direction from the source to the observer ($SA'$); the observable cascade $\gamma$ ray ($\gamma'$) is produced at $B'$ and intersects the observer's sphere (the sphere with the radius equal to the distance between the source $S$ and the observer $O$) at $O'$. Then the following check for the angle between $SO'$ and $SO$ ($\theta_{\mathrm{shift}}$) is performed:
\begin{equation}
    \theta_{\mathrm{shift}} \leq \theta_{\mathrm{jet}}.
    \label{neq:jet}
\end{equation}
If the condition (\ref{neq:jet}) is satisfied, then the polygonal line $SA'B'O'$ could be rotated around the pole $S$: $SA'B'O' \rightarrow SABO$ with $\gamma'$ ray reaching the Earth at $O$. If the condition (\ref{neq:jet}) is not satisfied, $\gamma'$ cannot reach the Earth, since this would require the source to emit primary $\gamma$ rays at angles greater than its jet half-opening angle $\theta_{\mathrm{jet}}$.

Furthermore, we have re-normalized $x$,$y$,$z$-components of the turbulent magnetic field $\mathbf{B}$ to ensure that the average squared values of all of them are equal, i.e.
\begin{equation}
\sqrt{\langle B_{x}^{2} \rangle} = \sqrt{\langle B_{y}^{2} \rangle} = \sqrt{\langle B_{z}^{2} \rangle} = \frac{B}{\sqrt{3}},
\end{equation}
using the \texttt{ELMAG 3} built-in \texttt{test\_turbB} function and following relevant instructions presented by \citet{Kalashev2022}. For every primary $\gamma$ ray we write the value of its energy $E_0$ to an output file; for every observable $\gamma$ ray we likewise write to the file the values of the following parameters: the energy $E$, the angle of the observable $\gamma$-ray w.r.t. the direction to the source $\theta = \sqrt{\theta_x^2 + \theta_y^2}$ (i.e. the angle between the arrival direction of the observable $\gamma$ ray and the source-observer line, this angle is shown in Fig. \ref{fig:scheme}), and the time delay $t_{d}$. An observable $\gamma$ ray may represent a cascade $\gamma$ ray or a primary redshifted $\gamma$ ray that did not absorb on the EBL or CMB (in the latter case $\theta = 0$). In order to ensure the correctness of our simulations, we reproduced Fig. 8 of \citet{Kalashev2022} obtained with the \texttt{CRbeam} code.

We performed a series of simulations of intergalactic electromagnetic cascades for the isotropic random non-helical turbulent magnetic field with a Kolmogorov spectrum with a wide range of the EGMF strength from $10^{-19}$ G to $10^{-12}$ G. The main input parameters of the \texttt{ELMAG 3} code for these simulations are shown in Table~\ref{tab:ELMAG3}. The values of all other parameters were set to their default values (they can be found in \texttt{ELMAG 3} \texttt{input\_b}, \texttt{input\_oth} and \texttt{input\_src} files in our Supplementary material available in Zenodo at \url{https://doi.org/10.5281/zenodo.6483355} \citep{Podlesnyi2022}).

\begin{table*}
\centering
\begin{tabular}{l|c}
Parameter & Value\\
\hline
Source cosmological redshift &  $0.186$\\
EBL model & \citet{Gilmore2012}\\
Minimal injection energy $E_{\mathrm{MIN}}$, eV & $10^{8}$\\
Maximal injection energy $E_{\mathrm{MAX}}$, eV & $10^{14}$\\
Power law spectral index before the break & 1.0\\
Power law spectral index after the break & 1.0\\
Jet opening angle $\theta_{\mathrm{jet}}$ in degrees & 6.0\\
Jet misalignment angle $\theta_{\mathrm{jetx}}$ in degrees & 0.0\\
Total number of injected primary $\gamma$ rays $n_{\mathrm{max}}$ & 60\,000\\
Number of turbulent modes $n_k$ & 200\\
EGMF minimal spatial scale, Mpc & $5 \times 10^{-4}$\\
EGMF maximal spatial scale, Mpc & $5$\\
EGMF correlation length $\lambda$, Mpc & 1.0\\
EGMF root mean square strength B, G & $\left\{ 10^{-19}, 10^{-18}, 10^{-17}, 10^{-16}, 10^{-15}, 10^{-14}, 10^{-13}, 10^{-12} \right\}$
\end{tabular}
\caption{Main \texttt{ELMAG 3} input parameters used in our simulations of extragalactic $\gamma$-ray propagation.}
\label{tab:ELMAG3}
\end{table*}

Having obtained the simulation results for each of the eight considered EGMF strength values, we constructed the corresponding eight three-dimensional arrays $A_{ijk}(E_{i}, E_{0_{j}}, \theta_{k}; B)$ containing the number of observable $\gamma$ rays with the observable energy $E$ falling into the $i$-th bin; the energy of the primary $\gamma$ ray $E_0$ falling into the $j$-th bin, and the line-of-sight deflection angle $\theta$ falling into the $k$-th bin, where $E_{i}$ is the central value of the $i$-th bin of the observable energy $E$, $E_{0_{j}}$ is the central value of the $j$-th bin of the primary $\gamma$-ray energy $E_0$, $\theta_{k}$ is the central value of the $k$-th bin of the observable deflection angle $\theta$.
        
We construct the following two-dimensional array:
\begin{equation}
A'_{ij}(E_{i}, E_{0_{j}}; B) = \sum \limits_{k\, : \, \theta_{k} \leq \theta_{68\%}(E_{i})} A_{ijk}(E_{i}, E_{0_{j}}, \theta_{k}; B),
\end{equation}
where the summation was applied over the cells of $A_{ijk}(B)$ with $\theta_{k} \leq \theta_{68\%}(E_{i})$, i.e those cells of the latter array that have the value of $\theta$ below or equal to the $68\%$-containment angle $\theta_{68\%}(E)$ of the \textit{Fermi}-LAT point-spread function (PSF)\footnote{Pass 8 Release 3 Version 2 Instrument Response Functions, \url{https://www.slac.stanford.edu/exp/glast/groups/canda/archive/pass8r3v2/lat_Performance.htm}} (or $0.1^{\circ}$ for the case of IACT observations).

The array $A'_{ij}$ now incorporates the \textit{Fermi}-LAT (or IACT) PSF effects. At the next step of our calculation, we obtain the observable $\gamma$-ray spectrum inside the $68\%$-containment angle of the PSF. We consider the following options for the intrinsic $\gamma$-ray spectrum of the source:
    \begin{enumerate}
        \item Power-law (PL):
        \begin{equation}
            \frac{dN}{dE_0} = C_0 \left( \frac{E_0}{E_{0_{\mathrm{ref}}}} \right)^{-\gamma_0}
            \label{eq:PL}
        \end{equation}
        \item Power-law with exponential cutoff (PLExp):
        \begin{equation}
            \frac{dN}{dE_0} = C_0 \left( \frac{E_0}{E_{0_{\mathrm{ref}}}} \right)^{-\gamma_0} \exp \left(- \frac{E_0}{E_{0_{\mathrm{cut}}}} \right)
            \label{eq:PLExp}
        \end{equation}
        \item Log parabola (LP):
        \begin{equation}
            \frac{dN}{dE_0} = C_0 \left( \frac{E_0}{E_{0_{\mathrm{ref}}}} \right)^{-[\alpha_0 + \beta_0 \ln (E_0 / E_{0_{\mathrm{ref}}})]}
            \label{eq:LP}
        \end{equation}
        \item Log parabola with exponential cutoff (LPExp):
        \begin{equation}
            \frac{dN}{dE_0} = C_0 \left( \frac{E_0}{E_{0_{\mathrm{ref}}}} \right)^{-[\alpha_0 + \beta_0 \ln (E_0 / E_{0_{\mathrm{ref}}})]} \exp \left(- \frac{E_0}{E_{0_{\mathrm{cut}}}} \right)
            \label{eq:LPExp}
        \end{equation}
    \end{enumerate}
The value of the reference energy was fixed ($E_{0_{\mathrm{ref}}} = 0.01$~TeV) during the whole analysis performed in the present paper. We consider only spectra with $\beta_{0} \geq 0$, $\gamma_{0} \geq 0.5$, and $\alpha_{0} \geq 0.5$.

The \texttt{ELMAG 3} simulations described above were obtained for the intrinsic power-law $\gamma$-ray spectrum $\propto E_{0}^{-1}$. To obtain the observable $\gamma$-ray spectrum for other shapes of the intrinsic $\gamma$-ray spectrum, we perform re-weighting with the weight $W(E_{0}; \mathbf{p})$ defined as the ratio of the new to the old intensity (see eq.~(13) of \citet{Dzhatdoev2017a}):
\begin{equation}
    \frac{dN}{dE_{i}} (E_{i}; B; \mathbf{p}) = \sum \limits_{j} A'_{ij}(E_{i}, E_{0_{j}}; B) W\left(E_{0_{j}}; \mathbf{p}\right),
    \label{eq:obs_spec}
\end{equation}
where $\mathbf{p}$ is the parameter vector of the corresponding spectral law (Eq. (\ref{eq:PL}), (\ref{eq:PLExp}), (\ref{eq:LP}), or (\ref{eq:LPExp})).

The observable spectrum strongly depends on the intrinsic spectrum parameters $\mathbf{p}$ and the EGMF strength $B$. In what follows, we determine which values of these parameters allow to describe the observational data well. Moreover, we exclude some values of $B$ that do not allow to fit the model SEDs to the observed SEDs.

\section{Statistical analysis} \label{sec:statistics}
\begin{figure*}
    \includegraphics[width=0.95\textwidth]{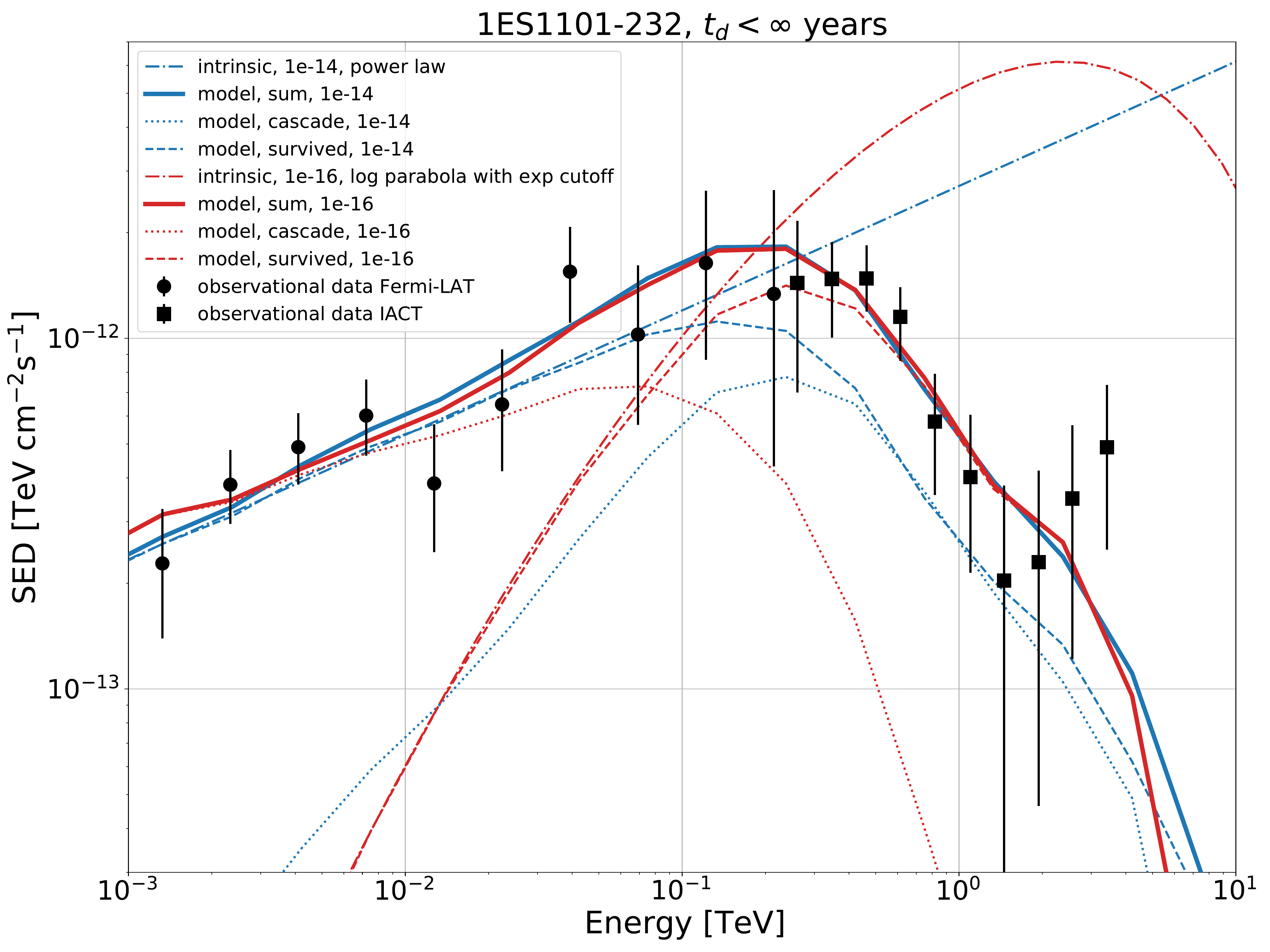}
    \caption{Example of the fitted SED of 1ES 1101-232 for two cases: i) PL, $B = 10^{-14}$~G and ii) LPExp, $B = 10^{-16}$~G. See legend and text for more details.  \label{fig:sed_example}}
\end{figure*}
\begin{figure*}
    \includegraphics[width=0.99\textwidth]{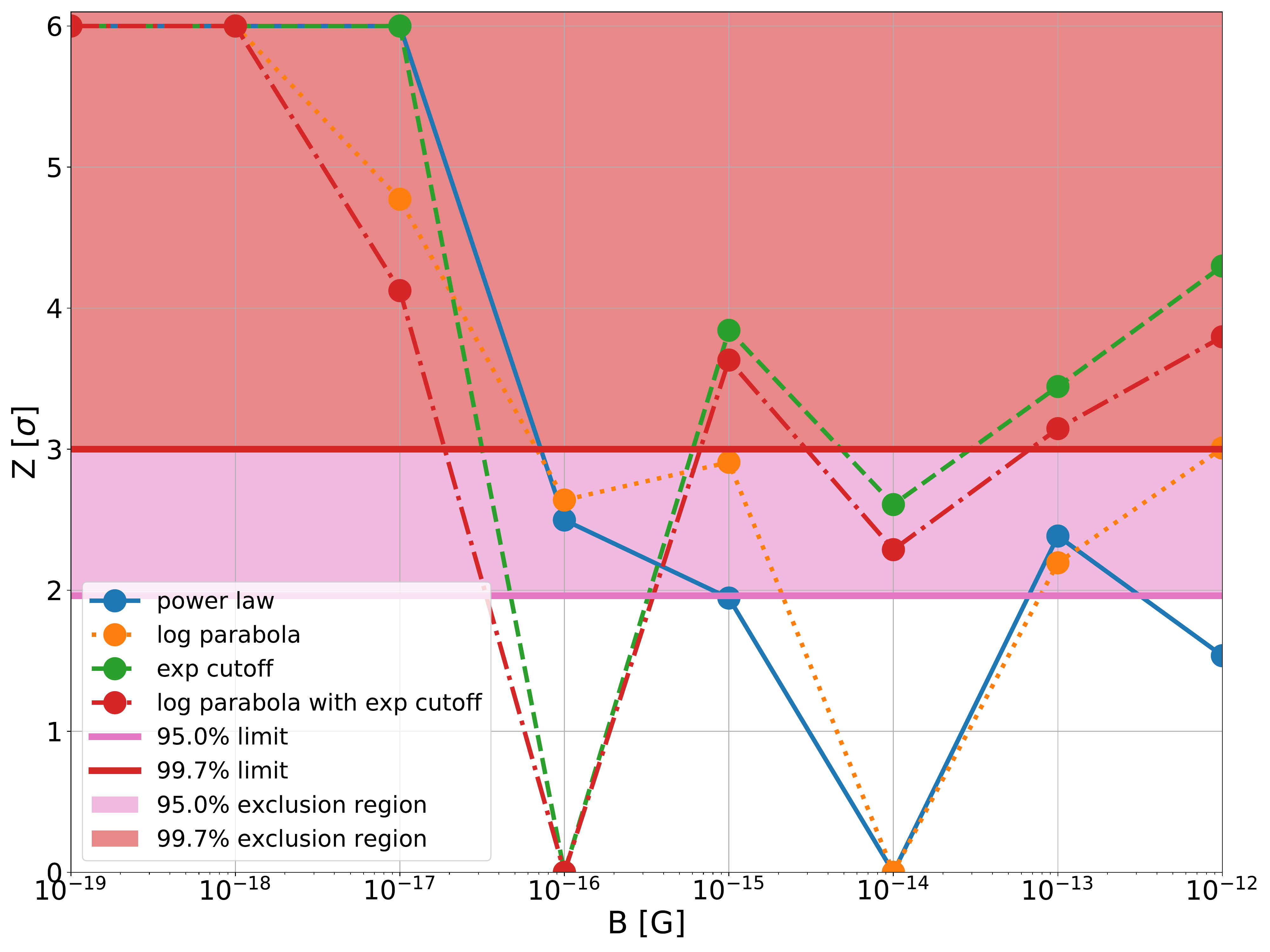}
    \caption{Exclusion empirical statistical significance $Z(B)$ for various intrinsic spectral shapes vs. the EGMF strength $B$ (see legend). The $95\%$ confidence level exclusion region is shown as pale pink area, $99.7\%$ confidence level exclusion region --- as pale red area. Values of $Z$ greater than 6 are shown as equal to 6. No constraints on $\gamma$-ray time delay $t_{d}$ are applied. \label{fig:results}}
\end{figure*}

We use the observed spectral energy distributions $\mathrm{SED}_{\mathrm{obs}} (E_{\mathrm{obs}_i})$ obtained with \textit{Fermi}-LAT and IACTs (see Sect. \ref{sec:Fermi-LAT}) to infer constraints on the EGMF strength $B$. We treat the intrinsic spectrum parameter vectors $\mathbf{p}$ as nuisance parameters and minimize the following functional form using the \texttt{scipy.optimize.least\_squares} \texttt{trf} method \citep{Branch1999}:
    \begin{equation}
        \chi^2(B; \mathbf{p}; \mathrm{\texttt{src}}) = \sum \limits_{i = 1}^{N_{\mathrm{obs}}} \frac{\left(\mathrm{SED}_{\mathrm{mod}} (E_{\mathrm{obs}_i}; B; \mathbf{p}) - \mathrm{SED}_{\mathrm{obs}} (E_{\mathrm{obs}_i})\right)^2}{\sigma^2_{\mathrm{obs}}(E_{\mathrm{obs}_i})},
        \label{eq:chi_sq}
    \end{equation}
where $$\mathrm{SED}_{\mathrm{mod}} (E_{\mathrm{obs}_i}; B; \mathbf{p}) \equiv E_{\mathrm{obs}_i}^{2} \frac{dN}{dE_{\mathrm{obs}_i}} (E_{\mathrm{obs}_i}; B; \mathbf{p}),$$
\texttt{src} represents the specific source, ${dN} / {dE_{\mathrm{obs}_i}} (E_{\mathrm{obs}_i}; B; \mathbf{p})$ is defined by interpolation of Eq. (\ref{eq:obs_spec}) from the nearest model bin with the central energy $E_{i}$ to the observational bin with the central energy $E_{\mathrm{obs}_i}$, $N_{\mathrm{obs}}$ is the total number of observational bins for a given $\gamma$-ray source, $\sigma_{\mathrm{obs}}(E_{\mathrm{obs}_i})$ is the measurement uncertainty of the observed SED in the $i$-th energy bin. For the case of \textit{Fermi}-LAT spectra, we account for only the statistical uncertainties. For some IACT measurements, the systematic uncertainties are available; if this is the case, $\sigma_{\mathrm{obs}}(E_{\mathrm{obs}_i})$ is calculated as $\sqrt{\sigma_{\mathrm{stat}}^{2}+\sigma_{\mathrm{syst}}^{2}}$, where $\sigma_{\mathrm{stat}}$ is the statistical uncertainty term and $\sigma_{\mathrm{syst}}$ is the systematic uncertainty term. We note that the statistical fluctuations of model SEDs are much lower than the observed SED measurement uncertainties. We did not perform any search for the variability of the considered sources and use observational data from the IACT observational periods stated in Table \ref{tab:sources} assuming they are compatible with the \textit{Fermi}-LAT data, since all the SED measurements have been conducted for at least several months and a hypothetical effect of possible flares is diminished by the time-averaging.

We perform a stacking statistical analysis for the set of the sources listed in Table~\ref{tab:sources} as follows. We calculate the stacked chi-square value
\begin{equation}
    \chi^2 (B; \hat{\mathbf{P}}(B)) = \sum \limits_{\texttt{src}} \chi^2 (B; \hat{\mathbf{p}}(B; \mathrm{\texttt{src}}))\
    \label{eq:chi_sq_stacked}
\end{equation}
for every set of the intrinsic spectrum parameters and every value of the EGMF strength $B$. Here $\hat{\mathbf{p}}(B; \texttt{src})$ is the value of $\mathbf{p}$ for the specific intrinsic spectrum yielding the minimum of the functional form (\ref{eq:chi_sq}) under the hypothesis of the EGMF strength B for the specific source \texttt{src}; $\hat{\mathbf{P}}(B)$ is the set of the best-fit values of vectors $\hat{\mathbf{p}}(B; \texttt{src})$ for all the considered sources.
    
Following an approach presented, e.g., by \citet{Meyer2016}, A18 we utilize the Wilks' theorem \citep{Wilks1938} and calculate the test statistic $TS$ as follows:
    \begin{equation}
        TS(B) = \chi^2 (B; \hat{\mathbf{P}}(B)) - \min\limits_{B} [\chi^2 (B; \hat{\mathbf{P}}(B))].
        \label{eq:TS}
    \end{equation}
The EGMF strength $B$ is the only parameter of interest in our analysis. Therefore, the test statistic (\ref{eq:TS}) is asymptotically distributed as the chi-square distribution with one degree of freedom. Finally, we calculate the empirical exclusion statistical significance $Z(B)$ for a two-tailed standard normal distribution corresponding to the value of $TS(B)$ separately for every option of the intrinsic spectrum (Eq. (\ref{eq:PL}), (\ref{eq:PLExp}), (\ref{eq:LP}), or (\ref{eq:LPExp})). The obtained results are discussed in Sect.~\ref{sec:results}.

\section{Results} \label{sec:results}

\subsection{An example of a fit to an observed SED}
\label{subsec:SED_example}
    
Before presenting constraints on the EGMF strength, let us show an example of an observed SED for the case of the blazar 1ES~1101-232 fitted with the model SEDs calculated according to eq.~(\ref{eq:obs_spec}) assuming two different values of $B$ (see Fig. \ref{fig:sed_example}) (examples of the fitted SEDs for 1ES 0347-121 and 1ES 1218+304 can be found in Appendix \ref{app:SED_examples}). Naturally, the best-fit intrinsic spectra for these two models are also significantly different. We note that the model SEDs for these two sets of substantially different parameters (namely: (i) a PL intrinsic spectrum for $B = 10^{-14}$~G and (ii) a LPExp intrinsic spectrum for $B = 10^{-16}$~G) result in approximately the same model SEDs. In the energy range of 1--30 GeV and case i) the model intensity is dominated by ``survived'' $\gamma$ rays (primary $\gamma$ rays that did not absorb on the EBL). On the contrary, in case ii) the model intensity in the same energy range is dominated by the cascade component, and the intensity of the latter is much greater than the intensity of the survived component.

Moreover, the assumption of A18 that ``accounting for the cascade contribution does not change the best-fit spectrum of the central point source in the entire \textit{Fermi}-LAT energy band by more than $5\sigma$'' appears to be not justified in case ii). In addition, A18 neglected any contribution of the cascade component to the SEDs measured with IACTs. This second assumption is also not always justified: for instance, in case i) the contributions of the cascade and survived components to the observable SED are comparable at the energy in excess of $300$~GeV.

\subsection{Exclusion significance vs. the EGMF strength}

The exclusion statistical significance $Z$ vs. the EGMF strength $B$ is shown in Fig.~\ref{fig:results}. The values of $B \le 10^{-17}$ G are excluded at a high level of significance $Z > 4 \sigma$ for all the four options of the intrinsic spectral shape. The best agreement of the model with the observed spectra for the case of the PLExp and LPExp intrinsic spectral shapes is achieved for $B = 10^{-16}$ G, and for the case of the PL and LP intrinsic spectral shapes --- for $B = 10^{-14}$ G. We note that $Z$ significantly depends on the assumed intrinsic spectral shape option.

Our results are significantly different from those presented by A18 (see their Fig.~17, right panel, $t_{\mathrm{max}} = 10^7$~years). The difference between their and our results is likely due to different assumptions. Namely, we do not assume that the contribution of the cascade component either to the \textit{Fermi}-LAT spectrum or to the IACT spectrum is small or subdominant (see the previous Subsection). Therefore, both options of $B = 10^{-16}$ G and $B = 10^{-14}$ G appear to be viable in our analysis.

\section{Discussion} \label{sec:discussion}

\subsection{The shape of the intrinsic spectrum}

The considered sources belong to the class of extreme TeV blazars (ETBs) (e.g. \citet{Biteau2020,Dzhatdoev2021}), i.e. the peak in their intrinsic SEDs is situated at an energy in excess of 1 TeV. The precise emission mechanism in ETBs is unknown and the knowledge on their intrinsic spectral shapes is limited. The assumed spectral shape could have some impact on the results of this work. We leave a more detailed study of this issue for future research.

\subsection{Possible influence of the plasma energy losses} \label{subsec:discussion:plasma}

Pair beams resulting from the development of intergalactic electromagnetic cascades may be subject to plasma instabilities that may cause additional energy losses w.r.t. the IC energy losses \citep[see, e.g.,][]{Broderick2012,Schlickeiser2012,Schlickeiser2013,Miniati2013,Chang2014,Sironi2014,Menzler2015,Kempf2016,Vafin2018,Vafin2019,Perry2021}. The impact of these ``plasma losses'' on the observable spectrum of the cascade component is, at present, unclear. Therefore, we have neglected plasma losses in the course of this work.

\subsection{Other systematic effects}

Constraints on the EGMF strength obtained from \textit{Fermi}-LAT and IACT measurements of blazar $\gamma$-ray spectra are subject to many systematic effects, including the following: 1) the uncertainty of the EBL models, 2) unknown beaming pattern of the blazars, uncertainties of their jet opening and viewing angles, 3) the unknown duty cycle of the sources, 4) the uncertainty of the void filling factor (``voidiness''), 5) a possible contribution of cascades initiated by ultrahigh energy cosmic rays \citep{Waxman1996,Uryson1998,Khalikov2021}. We note that the inclusion of any systematic effect is similar to the addition of a new nuisance parameter, leading to a further increase of the uncertainty of the EGMF strength constraints.

\subsection{Dependence on the correlation length}

We note that the results presented above were obtained for the EGMF correlation length $\lambda = 1$ Mpc. For a relatively small-scale EGMF with $\lambda < L_{E-e}$ (where $L_{E-e}$ is the electron energy loss length), the constraint on $B \propto \sqrt{L_{E-e}/\lambda}$ (e.g., \cite{Neronov2009}).

\subsection{Prospects of the EGMF constraints/measurements}

Next-generation $\gamma$-ray observatories with better sensitivity and angular resolution could allow to improve the limits on the EGMF strength significantly. For instance, the CTA IACT array \citep{Actis2011,Acharya2013} could significantly improve the intrinsic spectrum measurement accuracy, while next-generation space $\gamma$-ray telescopes such as MAST \citep{Dzhatdoev2019} could dramatically improve the cascade echo detectability prospects.
    
\section{Conclusions} \label{sec:conclusions}

In the present work we have obtained new constraints on the EGMF strength using 145 months of \textit{Fermi}-LAT observation of the blazars 1ES 1218+304, 1ES 1101-232, and 1ES 0347-121, as well as IACT observations of the same sources. We find that the values of the EGMF strength $B \le 10^{-17}$ G are excluded at a high level of statistical significance $Z > 4 \sigma$ for all the four options of the intrinsic spectral shape (power-law, power-law with exponential cutoff, log-parabola, log-parabola with exponential cutoff). On the other hand, $B = 10^{-16}$ G is still a viable value of the EGMF strength. These constraints were obtained for the case of steady sources. Next-generation $\gamma$-ray observatories such as CTA and MAST would allow to dramatically improve the sensitivity of $\gamma$-ray instruments for the cascade component in blazar spectra.

\section*{Acknowledgements}
We acknowledge helpful discussions with Prof. S. V. Troitsky.
This work made use of the following software: \texttt{ELMAG 3.01} \citep{Blytt2020}, \texttt{scipy} \citep{Virtanen2020}, \texttt{matplotlib} \citep{Hunter2007}, \texttt{numpy} \citep{vanderWalt2011, Harris2020}, \texttt{snakemake} \citep{Molder2021}, \texttt{fermipy} \citep{Wood2017}.
This research was partly funded by the Interdisciplinary Scientific and Educational School of Lomonosov Moscow State University ``Fundamental and Applied Space Research''. The work of E.P. and T.D. on the VHE $\gamma$-ray propagation in the intergalactic medium was supported by the Russian Science Foundation, grant No. 22-12-00253. E.P. thanks the Theoretical Physics and Mathematics Advancement Foundation ``BASIS'' (Contract No. 20-2-10-7-1) for the student scholarship.

\section*{Data Availability}
 
 The data used for the \textit{Fermi}-LAT data analysis in this work are publicly available at \url{https://fermi.gsfc.nasa.gov/cgi-bin/ssc/LAT/LATDataQuery.cgi}. The simulation data and software underlying the main article results are available in Zenodo at \url{https://doi.org/10.5281/zenodo.6483355} \citep{Podlesnyi2022}.

\bibliographystyle{mnras}
\bibliography{bibliography}



\appendix

\section{Examples of the fitted SEDs of 1ES 0347-121 and 1ES 1218+304}
\label{app:SED_examples}

\begin{figure*}
    \includegraphics[width=0.75\textwidth]{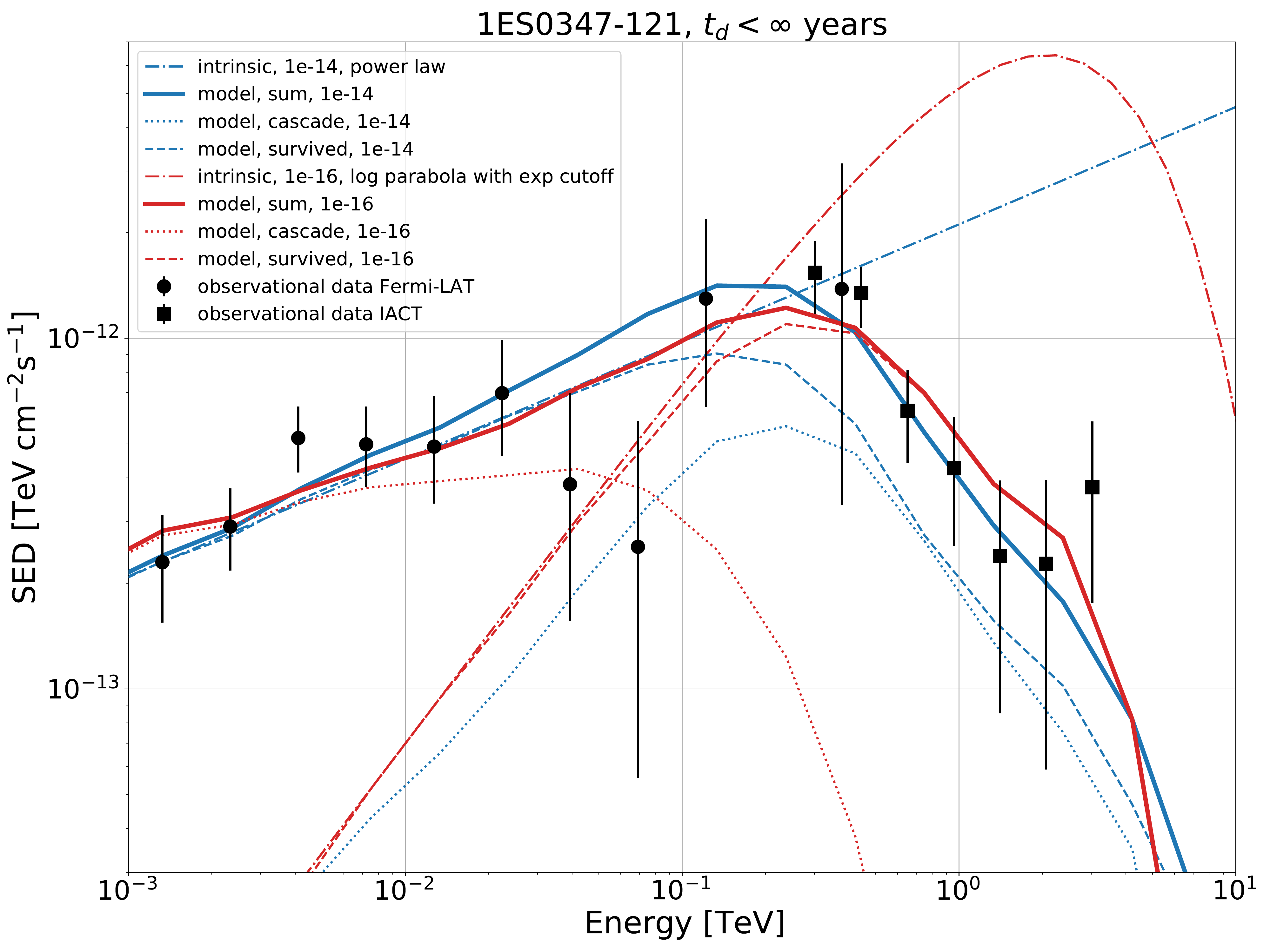}
    \caption{Example of the fitted SED of 1ES 0347-121 for two cases: i) PL, $B = 10^{-14}$~G and ii) LPExp, $B = 10^{-16}$~G. \label{fig:SED_1ES0347-121}}
\end{figure*}

\begin{figure*}
    \includegraphics[width=0.75\textwidth]{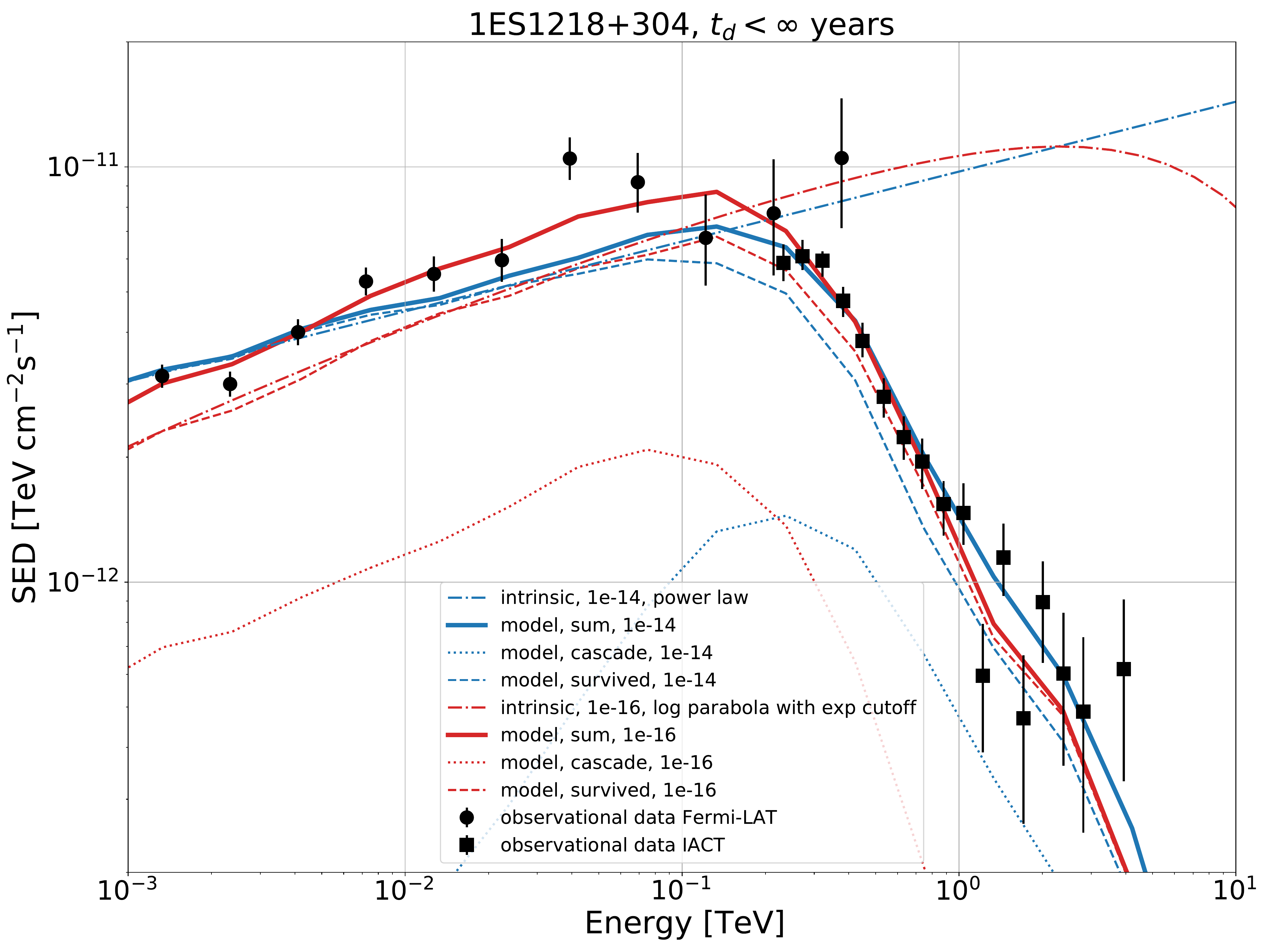}
    \caption{Example of the fitted SED of 1ES 1218+304 for two cases: i) PL, $B = 10^{-14}$~G and ii) LPExp, $B = 10^{-16}$~G. \label{fig:SED_1ES1218+304}}
\end{figure*}

Here we present examples of the fits to observed SEDs of 1ES 0347-121 (Fig.~ \ref{fig:SED_1ES0347-121}) and 1ES 1218+304 (Fig.~\ref{fig:SED_1ES1218+304}) for the same EGMF strengths and source intrinsic spectral shapes as in Sect. \ref{subsec:SED_example}.


\section{Individual contributions of the considered blazars to the $\chi^{2}$ functional}
\label{app:chisq_contributions}
\begin{figure*}
    \includegraphics[width=0.99\textwidth]{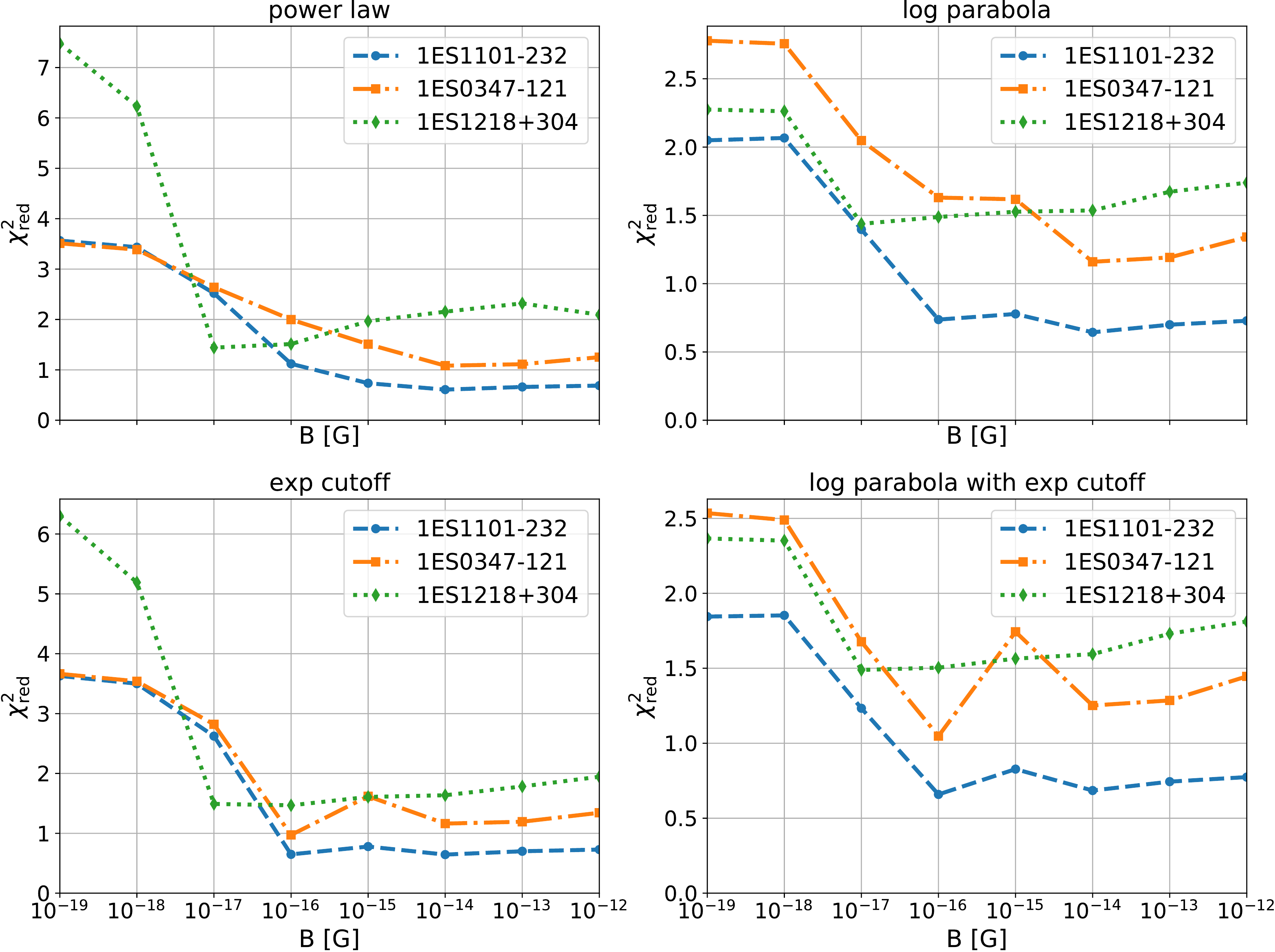}
    \caption{Values of the reduced chi-square functional form for each source and every intrinsic spectrum shape considered vs. the EGMF strength B. Upper left: PL, lower left: PLExp, upper right: LP, lower right: LPExp. \label{fig:chisq_contributions}}
\end{figure*}

In this Appendix we present the goodness-of-fit for individual sources vs. the EGMF strength. For this purpose we calculate the reduced chi-square functional form
\begin{equation}
    \chi^{2}_{\mathrm{red}}(B; \mathbf{p}; \mathrm{\texttt{src}}) = \frac{\chi^2(B; \mathbf{p}; \mathrm{\texttt{src}})}{\mathrm{n.d.o.f.}},
\end{equation}
where $\chi^2(B; \mathbf{p}; \mathrm{\texttt{src}})$ is defined by Eq. (\ref{eq:chi_sq}), $\mathrm{n.d.o.f.}$ is the number of degrees of freedoms, which equals the number of observational data points for the specific source $\mathrm{\texttt{src}}$ minus the number of free parameters of the considered intrinsic $\gamma$-ray spectrum (Eq. (\ref{eq:PL}), (\ref{eq:PLExp}), (\ref{eq:LP}), or (\ref{eq:LPExp})). The obtained values of $\chi^{2}_{\mathrm{red}}(B; \mathbf{p}; \mathrm{\texttt{src}})$ are presented in Fig. \ref{fig:chisq_contributions}. One can see that for the PL and PLExp intrinsic spectral shapes the exclusion significance of weak EGMF strengths is dominated by the 1ES 1218+304 contribution, whereas for the LP and LPExp cases the contributions of all the three blazars are comparable.


\bsp 
\label{lastpage}
\end{document}